\documentclass[10pt,dvips,a4paper]{article}
\usepackage{epsfig}

\include{commands}
\usepackage{amssymb}
\usepackage{amsmath}
\usepackage{setspace}
\usepackage[numbers]{natbib}
\usepackage{lineno}
\begin{document}

\renewcommand{\title}{A lattice approach to model flow in cracked concrete}

\begin{center} \begin{LARGE} \textbf{\title} \end{LARGE} \end{center}

\begin{center} 
Peter Grassl\\
Department of Civil Engineering, University of Glasgow, Glasgow, UK\\
grassl@civil.gla.ac.uk \vspace{4cm}\\

Accepted in Cement and Concrete Composites\\
Last revision 27th of April 2008
\end{center}

\section*{Abstract}
This paper presents a lattice approach to model the influence of cracking on inviscid flow in concrete.
A mechanical lattice model based on a damage-plasticity constitutive model was combined with a new dual lattice of conduit elements for flow analysis.
The diffusivity of the conduit elements depends on the crack-opening obtained from the mechanical lattice.
The coupled lattice model was applied to several benchmark tests for aligned and random lattices.
The results for mechanical loading and flow analysis obtained with the new approach were shown to be independent of the size of lattice elements used.\\
Keywords: Concrete, cracks, lattice, transport, flow, Voronoi tesselation

\section{Introduction}
Cracking increases the permeability and diffusivity of concrete, which may accelerate the deterioration of concrete structures \cite{WitWanIwaGal80,GerBreAmm96,WanJanSha97,HosBinNem09}.
Cracks can act, for instance, as pathways for water containing chlorides, which promote corrosion of reinforcement and subsequently may lead to further cracking and spalling of the concrete cover.
For cracking induced by mechanical loading, macroscopic crack widths in concrete are a result of complex fracture processes on the meso-scale. 
For instance, crack bridging and crack tortuosity depend on the material composition and may influence permeability.  
Other types of loading, such as drying shrinkage subjected to aggregate restraint, lead to distributed micro-cracking, which depends on aggregate size and volume fraction and cannot be uniquely related to a macroscopic crack width \cite{WonZobBue08, GraWonBue08}.
Numerical analysis of the fracture process of concrete on the meso-scale within the framework of the finite element method may increase the understanding of the influence of material composition on micro-cracking and transport properties.
However, this requires the development of robust and accurate analysis methods for the interaction of fracture and mass transport.

Modelling of flow in cracked concrete within the finite element framework can be categorised according to nonlinear fracture mechanics approaches.
In continuum approaches, fracture is represented as regular fields of localised inelastic strains of finite size by using higher order constitutive laws, such as integral-type nonlocal models \cite{BazPij88, BazJir02}. Transport processes in the continuum can be spatially varied according to the values of accumulated history variables, such as inelastic strains. This approach is computationally demanding since a fine discretisation is required to be able to describe the localised strain fields.
Alternatively, the representation of the fracture process zone is simplified to a displacement discontinuity embedded into the continuum \cite{MoeBel02, Jir00b, CamOrt96}.
For these hybrid approaches, localised transport along the discrete crack is coupled with continuum modelling of imbibition into the adjacent material \cite{RoeVanCar03,CarDelVanRoe04,SegCar04,RoeMooDeProCar06}. 
Coupling of the flow in discrete cracks and the surrounding material can create difficulties in these approaches \cite{RoeMooDeProCar06}.
Finally, in discrete approaches the fracture process is described by the failure of structural elements, such as trusses and beams. 
Lattice models have shown to be capable of describing complex fracture patterns on the meso-scale of concrete \cite{HerHanRou89, SchMie92b, DelPijRou96}.
Furthermore, mass transport can be described by a lattice of conduit elements, which can be linked to the structural lattice to couple fracture and transport processes \cite{ChaPicKhePij05}.
Some lattice approaches exhibit mesh dependency and are limited in describing the continuum response accurately.
A special type of lattice model for the mechanical response and mass transport was proposed recently, which provides mesh-independent and accurate descriptions of basic aspects of the continuum response  \cite{BolSai98, BolSuk05, BolBer04}.
In this approach, the cross-sections of structural and transport elements are determined from the Voronoi tessellation of random points placed in the strutural domain.
This modelling approach was further developed to describe the interaction of transport along discrete cracks and the surrounding material by introducing an additional lattice of transport elements \cite{NakSriYas06, WanSodUed08}. 
This research direction is analogous to the hybrid approaches, with the challenges arising in the interaction of the two lattices describing transport along cracks and transport in the adjacent material.

In the present study, a new lattice approach is developed to provide a mesh-independent description of the fracture process and mass transport in the cracked material.
Transport is modelled by one lattice, which is dual to the lattice used for the mechanical response.
The change of transport properties due to localised crack openings is smeared out over the width represented by the lattice element, ensuring a mesh-independent description.
The first objective of this work is to demonstrate that a lattice dual to the mechanical network accurately represents mass transport in the uncracked continuum.
The second objective is to investigate if flow in a cracked media, such as concrete, can be represented independently of the size of the element mesh.
To the author's knowledge, this is the first lattice approach that models flow in fractured media independently of the mesh size.

\section{Combined model for mechanical loading and flow}

\subsection{Mechanical model}
In this work the mechanical response of concrete is described by a lattice approach \cite{Kaw78,BolSai98} with a damage-plasticity constitutive model \cite{GraRem08}.
In the following sections the modelling approach is briefly reviewed.

\begin{figure}
\begin{center}
\begin{tabular}{cc}
\epsfig{file=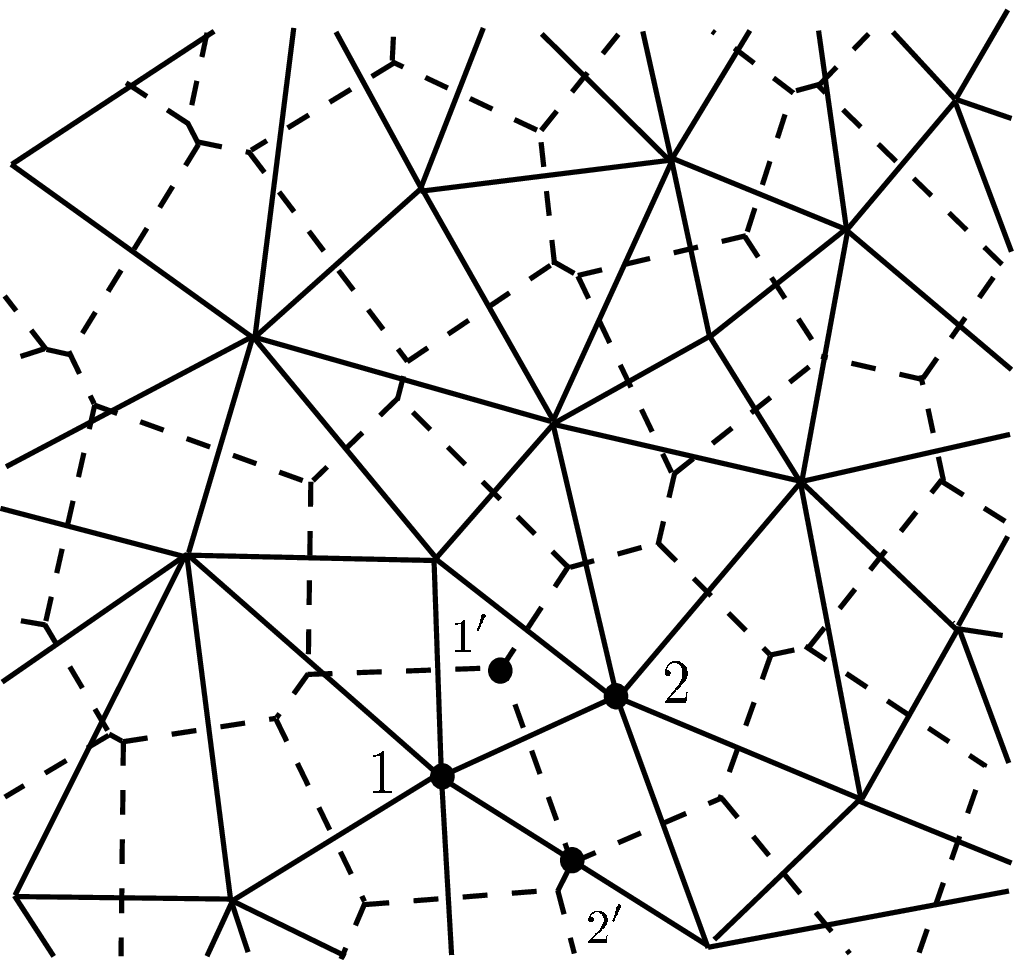,width=6.cm} & \epsfig{file=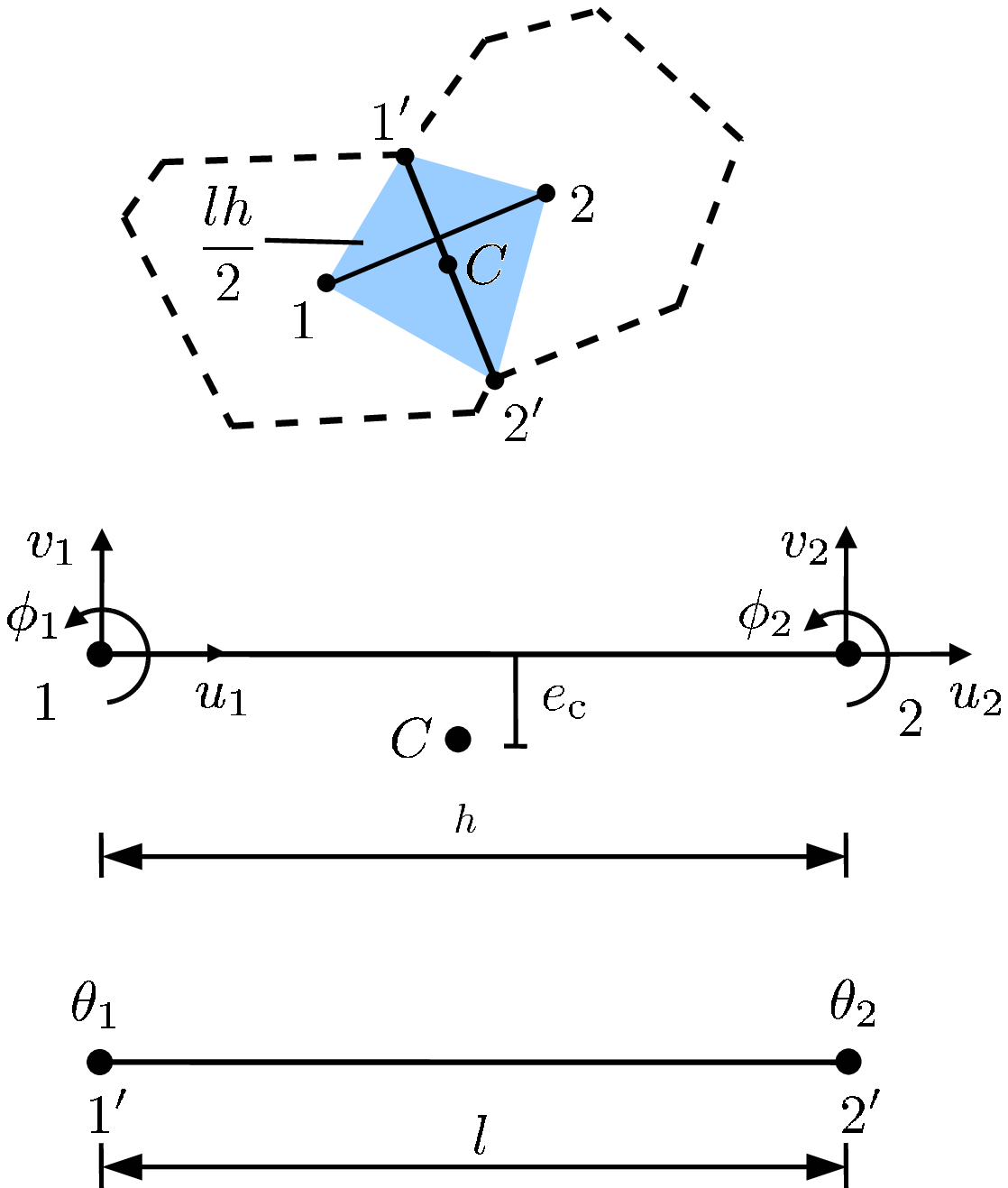,width=6cm} \\
(a) & (b)
\end{tabular}
\end{center}
\caption{Discretisation: (a) Elements (solid lines) and corresponding cross-sections (dashed lines) of the mechanical lattice obtained from the Delaunay triangulation and dual Voronoi tessellation, respectively. (b) Degrees of freedoms of the lattice elements for the mechanical and transport model in the local co-ordinate system.}
\label{fig:vorDel}
\end{figure}

The lattice model is based on the Voronoi tesselation of the domain \cite{OkaBooSug00}.
Lattice elements connect nodes, which are the nuclei of the neighboring Voronoi polygons (Fig.~\ref{fig:vorDel}a).
The nodes are placed randomly in the domain, constrained by a minimum allowable distance $d_{\rm m}$ between nodes.
The smaller $d_{\rm m}$, the smaller is the average distance between the nodes and the average size of the lattice elements \cite{BolSai98}.
Each node has three degrees of freedom (Fig.~\ref{fig:vorDel}b), that is two translations and one rotation, which determine the displacement jump at midpoint $C$ of the element cross-section in the local coordinate system as
\begin{equation}\label{eq:uc}
\mathbf{u}_{\rm c} = \mathbf{B} \mathbf{u}_{\rm e}
\end{equation}
where
\begin{equation}\label{eq:bMatrix}
\mathbf{B} = \begin{bmatrix}
-1 & 0 & e & 1 & 0 & -e\\
0 & -1 & -h/2 & 0 & 1 & -h/2
\end{bmatrix}  
\end{equation}
and $\mathbf{u}_{\rm e} = \left\{u_1, v_1, \phi_1, u_2, v_2, \phi_2\right\}^{\rm T}$.
In Eq.~(\ref{eq:bMatrix}) $e$ is the eccentricity of the midpoint $C$ with respect to the element axis and $h$ is the length of the element (Fig.~\ref{fig:vorDel}b). If $C$ is on the left hand side of the element, $e=e_{\rm c}$. Otherwise, $e = -e_{\rm c}$.

The cross-sections of the lattice elements are the faces of the Voronoi polygons.
The element stiffness in the local coordinate system is 
\begin{equation}
\mathbf{K}_{\rm e} = \dfrac{l}{h} \mathbf{B}^{\rm T} \mathbf{D}_{\rm e} \mathbf{B}
\end{equation}
where $l$ is the width of the cross-section (polygon facet) and $\mathbf{D}_{\rm e}$ represents the material properties as described below.
The present model was proposed for two-dimensional plane-stress analysis; for the examples considered later, the out-of-plane thickness was assumed to be equal to 1.

The constitutive model, which relates the strain vector $\boldsymbol{\varepsilon} = \mathbf{u}_{\rm c}/h$ to the nominal stress vector $\boldsymbol{\sigma}$, is based on a combination of plasticity formulated in the effective stress space and isotropic damage mechanics.
The stress-strain law is
\begin{equation} \label{eq:totStressStrain}
\boldsymbol{\sigma} = \left(1-\omega \right) \mathbf{D}_{\rm e} \left(\boldsymbol{\varepsilon} - \boldsymbol{\varepsilon}_{\rm p} - \boldsymbol{\varepsilon}_{\rm T}\right) = \left(1-\omega\right) \bar{\boldsymbol{\sigma}}
\end{equation}
where $\omega$ is the damage variable, $\mathbf{D}_{\rm e}$ is the elastic stiffness,  $\boldsymbol{\varepsilon}_{\rm p} = \left(\varepsilon_{\rm pn}, \varepsilon_{\rm ps}\right)^T$ is the plastic strain, $\boldsymbol{\varepsilon}_{\rm T} = \left(\varepsilon_{\rm T}, 0\right)^T$ is the eigenstrain, and $\bar{\boldsymbol{\sigma}} = \left(\bar{\sigma}_{\rm n}, \bar{\sigma}_{\rm s}\right)^T$ is the effective stress. 
The subscripts $n$ and $s$ denote the normal and shear direction in the local coordinate system (Fig.~\ref{fig:vorDel}b).
The eigenstrain is defined here as a strain which is not due to mechanical loading, but to other physical or chemical processes.
The elastic stiffness is 
\begin{equation}
\mathbf{D}_{\rm e} = \begin{Bmatrix} E & 0\\
  0 & \gamma E
\end{Bmatrix}
\end{equation}
where $E$ and $\gamma$ are model parameters controlling both the Young's modulus and Poisson's ratio of the material \cite{GriMus01}.
For instance, for plane stress considered here and a lattice of equilateral triangles, Poisson's ratio $\nu$ is
\begin{equation}
\nu = \dfrac{1-\gamma}{3+\gamma}
\end{equation}

The plasticity part of the damage-plasticity model is based on the effective stress and consists of the yield surface, flow rule, evolution law for the hardening parameter, and loading-unloading conditions.
The yield surface is elliptic and its initial size and shape are determined by the tensile strength $f_{\rm t}$, the shear strength $s f_{\rm t}$ and the compressive strength $c f_{\rm t}$ of the material under consideration.
The evolution of the yield surface during hardening is controlled by the model parameter $\mu$, which is defined as the ratio of permanent and total inelastic displacements.
The scalar damage part is chosen so that linear stress inelastic displacement laws for pure tension and compression are obtained, which are characterised by the fracture energies $G_{\rm ft}$ and $G_{\rm fc}$. 
The equivalent crack opening is defined as $\tilde{w}_{\rm c} = \|\mathbf{w}_{\rm c}\|$, where 
\begin{equation} \label{eq:crackOpening}
\mathbf{w}_{\rm c} = h \left( \boldsymbol{\varepsilon}_{\rm p} + \omega \left(\boldsymbol{\varepsilon} - \boldsymbol{\varepsilon}_{\rm p}\right)\right)
\end{equation}
The crack opening vector $\mathbf{w}_{\rm c}$ is composed of a permanent and reversible part, defined as $h \boldsymbol{\varepsilon}_{\rm p}$ and $\omega h \left(\boldsymbol{\varepsilon} - \boldsymbol{\varepsilon}_{\rm p}\right)$, respectively \cite{GraRem08}.

\begin{figure} [ht]
\begin{center}
\epsfig{file=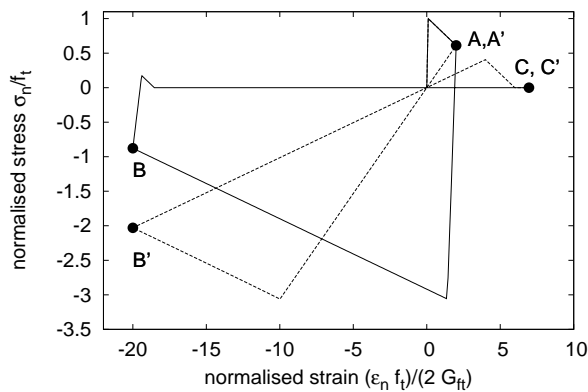, width=8cm}
\end{center}
\caption{Stress-strain response for fluctuating normal strains for  $\mu = 1$ (solid line) and $\mu = 0$ (dashed line).}
\label{fig:constCyclic}
\end{figure}

The constitutive behavior of the damage-plasticity model is illustrated by its stress-strain response for fluctuating normal strains for $\mu = 1$ and $\mu = 0$ (Figure~\ref{fig:constCyclic}).
The normal strain is increased to point~$A$ for the case of $\mu=1$ (and to point~$A^{'}$ for $\mu=0$).
$A$~and~$A^{'}$ coincide, since the response for both parameter settings has been identical up to this point.
The normal strain is then reduced to point~$B$~($B^{'}$) and again increased to point~$C$~($C^{'}$).
For $\mu = 1$ the combined approach reduces to a pure plasticity model.
The unloading is elastic and under subsequent compressive loading the compressive strength is reached sooner than for $\mu = 0$.
On the other hand, a pure damage-mechanics response is obtained for $\mu=0$. The stress-strain curve is unloaded to the origin.
This constitutive model in combination with the present lattice approach results in a mesh-independent description of the mechanical response.
A detailed description of the components of the model is presented in \cite{GraRem08}.

\subsection{Coupling of flow and mechanical models}\label{sec:transport}
Here, flow is modelled by a new lattice approach, for which the spatial arrangement of the conduit elements is dual to the arrangement of structural elements. 
The conduit elements are placed along the facets of the Voronoi polygons (Fig.~\ref{fig:vorDel}a).
Equivalently, cross-sections of conduit elements are positioned on the mechanical lattice elements. 
The discrete form of the differential equation of the non-stationary flow problem for one conduit element is defined as
\begin{equation} \label{eq:discreteFlow}
\boldsymbol{\alpha}_{\rm e} \boldsymbol{\theta} + \mathbf{C}_{\rm e} \dfrac{\partial \boldsymbol{\theta}}{\partial t}  = \mathbf{f}
\end{equation}
where $\boldsymbol{\alpha}_{\rm e}$ and $\mathbf{C}_{\rm e}$ are the element conductivity and capacity matrix, respectively, $t$ is the time, $\mathbf{f}$ are the external fluxes, and the degrees of freedom of the conduit elements are the flow potential $\boldsymbol{\theta} = \left(\theta_1, \theta_2\right)^T$ (Fig.~\ref{fig:vorDel}b) \cite{BolBer04}.
The conductivity matrix is defined as
\begin{equation}
\boldsymbol{\alpha}_{\rm e} = \dfrac{h}{l} \alpha
\begin{pmatrix} 
1 & -1\\
-1 & 1
\end{pmatrix}
\end{equation}
where $h$ is cross-section width, $l$ is the length of the pipe element and $\alpha$ is the conductivity of the material.
The capacity matrix $\mathbf{C}_{\rm e}$ is
\begin{equation}
\mathbf{C}_{\rm e} = \dfrac{h l}{12}
\begin{pmatrix}
2 & 1\\
1 & 2
\end{pmatrix}
\end{equation}

Mechanical loading is assumed to influence the diffusivity of the conduit elements as
\begin{equation} \label{eq:diffSplit}
\alpha = \alpha_0 + \alpha_{\rm c} \left( h \right)
\end{equation}
where $\alpha_0$ is the initial diffusivity of the undamaged material and $\alpha_{\rm c}$ is the change of diffusivity due to mechanical loading.
The part $\alpha_{\rm c}$ differs strongly depending on the problem modelled. For moisture transport, for instance, it could be related to the cubic law \cite{WitWanIwaGal80}. 
In the present study a simple linear law of the form
\begin{equation}\label{eq:crackDiff}
\alpha_{\rm c} = \alpha_0 \dfrac{\tilde{w}_{\rm c}} {h \varepsilon_{\rm fk}} 
\end{equation}
was chosen, where $\tilde{w}_{\rm{c}}$ is the equivalent crack opening from Eq.~(\ref{eq:crackOpening}) and $\varepsilon_{\rm fk}$ is a parameter which controls the slope of the change of diffusivity.
The equivalent crack opening is determined in Eq.~(\ref{eq:crackOpening}) from the structural lattice element which crosses the conduit element. 
Since the structural and transport lattices are dual, the cross-section width $h$ of the conduit element is equal to the length of the structural lattice element. 
An important aspect of the proposed model is the mesh independence, which is achieved by introducing the cross-section $h$ in the expression of the diffusivity in Eq.~(\ref{eq:crackDiff}), so that the change of conductivity due to mechanical loading in Eq.~(\ref{eq:discreteFlow}) is independent of the width of the conduit elements used.
The parameters $\alpha_0$ and $\varepsilon_{\rm fk}$ are material parameters.

\section{Model results}
The modelling approach described above, which was implemented in the object oriented finite element code OOFEM \cite{Pat99, PatBit01}, was applied to three benchmark problems. 
In the first example, a stationary flow field is represented on a random lattice. 
The second example involves the coupling of mechanical loading and flow for a lattice, which is aligned to a potential crack path. 
Finally, mechanical loading and flow are described for radial cracking about an inclusion within a random lattice.
The numerical results of the first example are compared to the analytical solution.
For the other two examples, possible mesh-dependence of the numerical results is investigated.

\subsection{Stationary flow within homogeneous media}
A graded lattice with conduit elements on the facets of Voronoi polygons is shown in Fig.~3.
\begin{figure}
\begin{center}
\epsfig{file=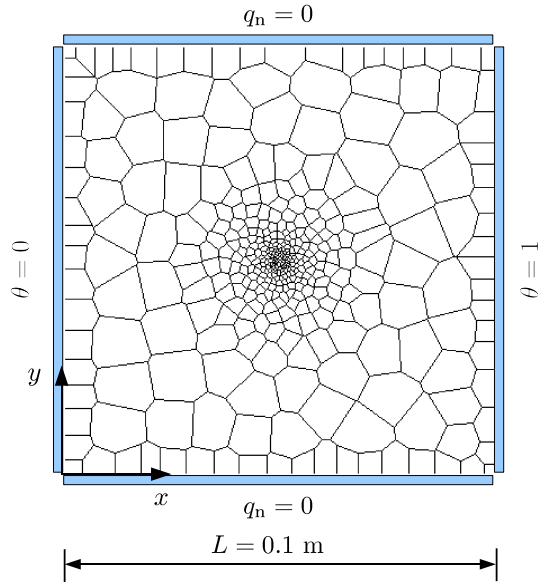, width=7cm}
\end{center}
\caption{Discretisation of the domain by means of a graded network. Lattice elements are located on the facets of the Voronoi polygons.}
\label{fig:meshGraded}
\end{figure}
The cross-sectional areas of the conduit elements were chosen in two ways.
In the first approach, cross-sectional areas were determined by the dual Delaunay triangulation as described in Section~\ref{sec:transport}.
In the second approach, a constant cross-sectional was used, which is the average of the areas obtained from the first approach.
The nodes on the left and right hand sides of the model domain of length $L = 0.1$~m were subjected to constant potentials of $\theta = 0$ and $\theta = 1$, respectively.
For the other two edges, the boundary flux was assumed to be zero ($q_{\rm n} = 0$).
The diffusivity was chosen as $\alpha=\alpha_0=1$~m$^2$/s.
The exact solution for this problem is $\theta=10x$.
The flow along the $x$-direction for $y=L/2$ for the two approaches is shown in Fig.~\ref{fig:xFlow1}.
\begin{figure}
\begin{center}
\epsfig{file = 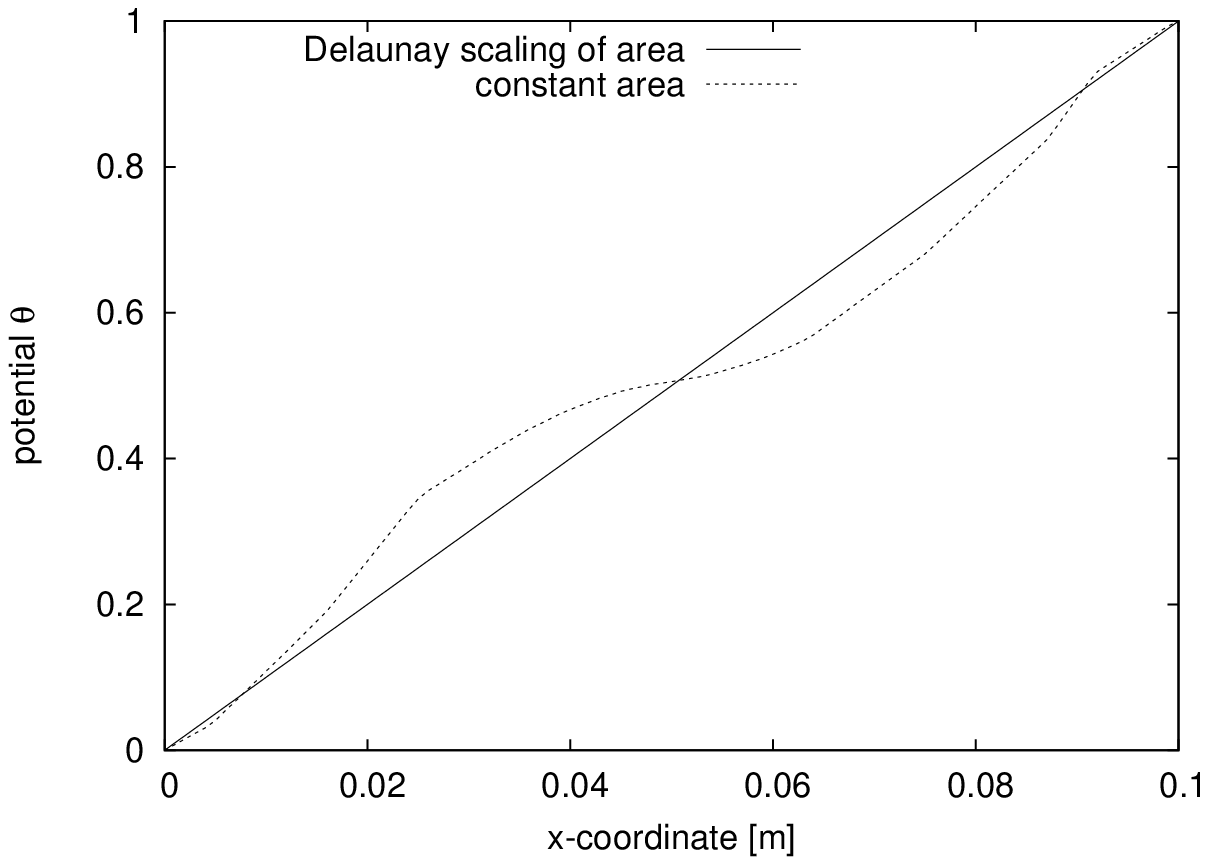, width=9cm}
\end{center}
\caption{Comparison of the potential $\theta$ along the $x$-direction ($y = L/2$) for cross-sections obtained with the Delaunay triangulation and a constant average cross-section.}  
\label{fig:xFlow1}
\end{figure}
Additionally, the accuracy of the modelling approach was assessed by comparing the $L_2$ error norm to the exact solution.
The error norm is
\begin{equation}
e_{\rm r} = \dfrac{\| \theta - \theta^{\rm h}\|_2}{\|\theta\|_2} 
\end{equation}
where
\begin{equation}
\|\theta - \theta^{\rm h}\|_2 = \sqrt{\sum^{N}_{I=1} \left(\theta\left(\mathbf{x}_I\right) - \theta^h\left(\mathbf{x}_I\right)\right)^2}
\end{equation}
and $\mathbf{x}$ is the position of node $I$, and $\theta$ and $\theta^h$ are the exact and numerical values of the potential, respectively.
Furthermore, $N$ is the number of nodes in the specimen.
The error for a constant cross-sectional area is $e_{\rm r} = 0.052$, whereas the error for cross-sectional areas obtained from the Delaunay triangulation is $e_{\rm r} = 5.795 \times 10^{-10}$. 

Consequently, the lattice with transport elements placed on the edges of Voronoi tesselation and cross-sections obtained from the dual Delaunay triangulation results in an accurate description of the stationary flow field. 
The results of the present study complement results obtained from a dual lattice approach, in which the conduit elements are placed on the edges of the Delaunay triangulation and the cross-sections are the facets of the Voronoi tessellation \cite{BolBer04}. 

\subsection{Nonstationary flow along a planar crack}
\begin{figure}
\begin{center}
\begin{tabular}{cc}
\epsfig{file=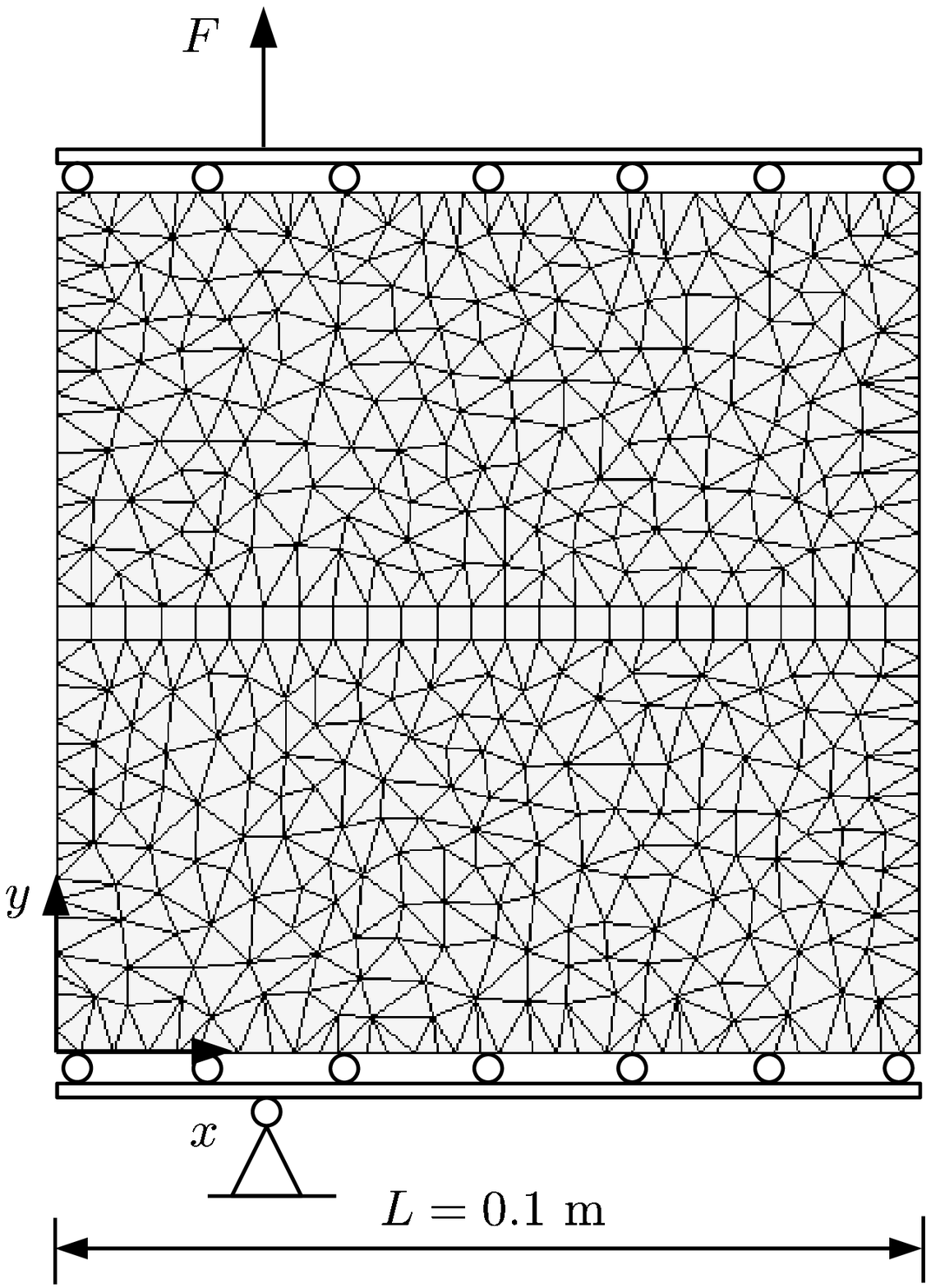,width = 6cm} & \epsfig{file=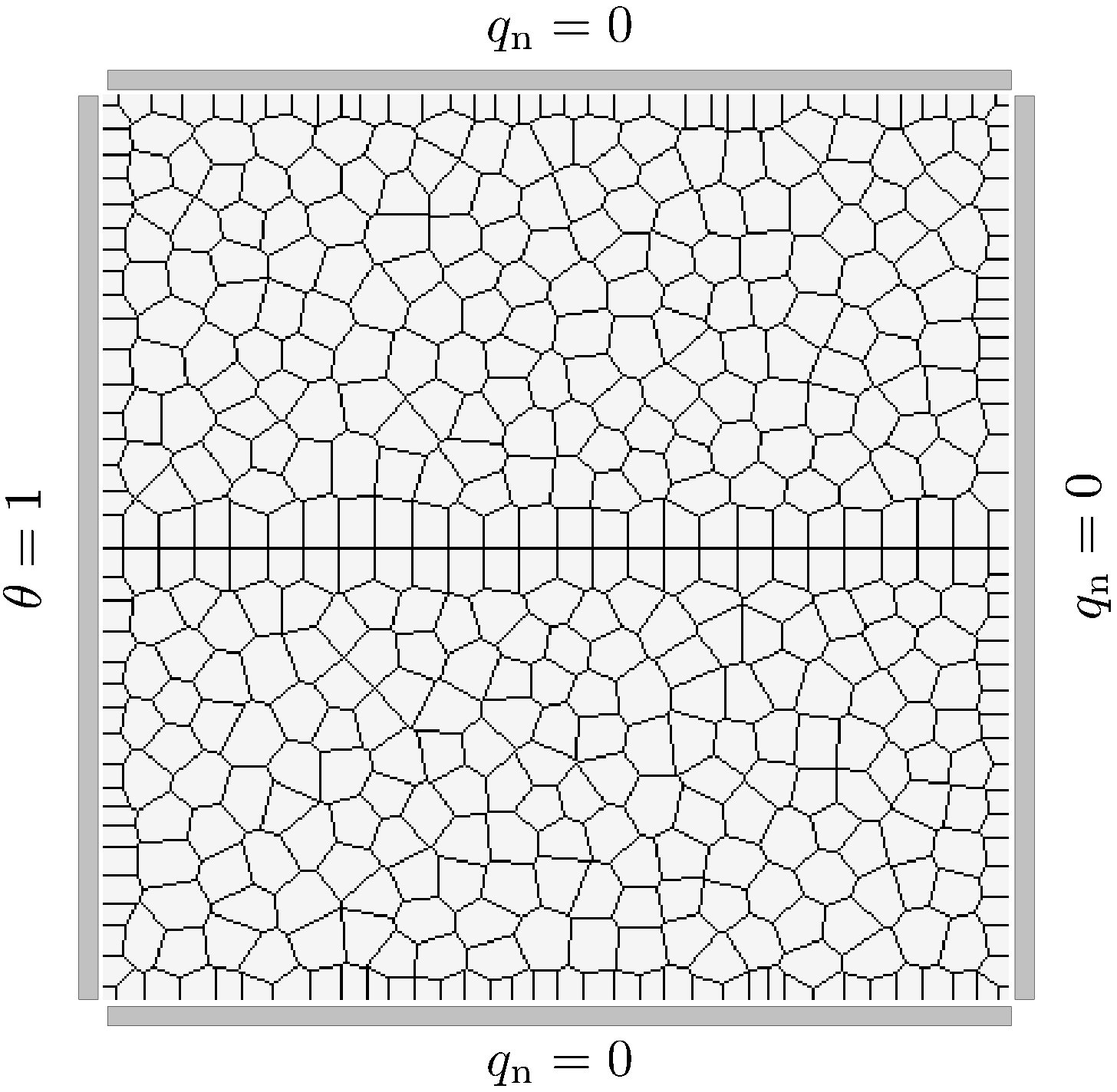,width = 6.8cm}\\
(a) & (b)
\end{tabular}
\end{center}
\caption{Meshes and loading set-up for (a) structural and (b) transport analysis for the medium density lattice with elements aligned along the potential crack path.}
\label{fig:mesh2}
\end{figure}

\begin{figure}
\begin{center}
\epsfig{file=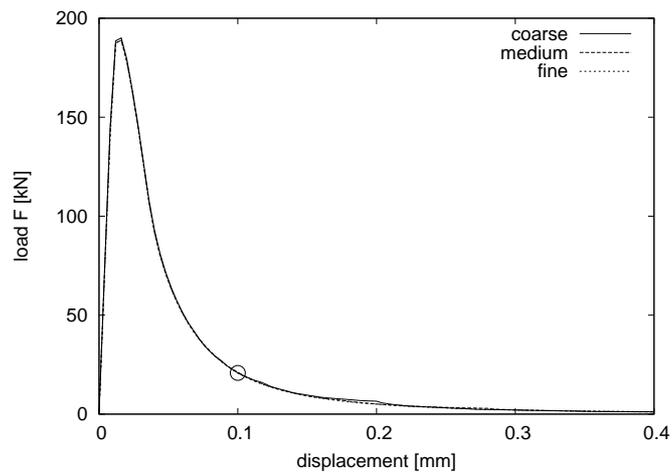,width = 9cm}
\end{center}
\caption{Results of the structural analysis: Load-displacement curves for three element sizes. The open circle marks the stage for which the flow analyses are performed in the second stage of the analysis.}
\label{fig:ld2}
\end{figure}

\begin{figure}
\begin{center}
\epsfig{file=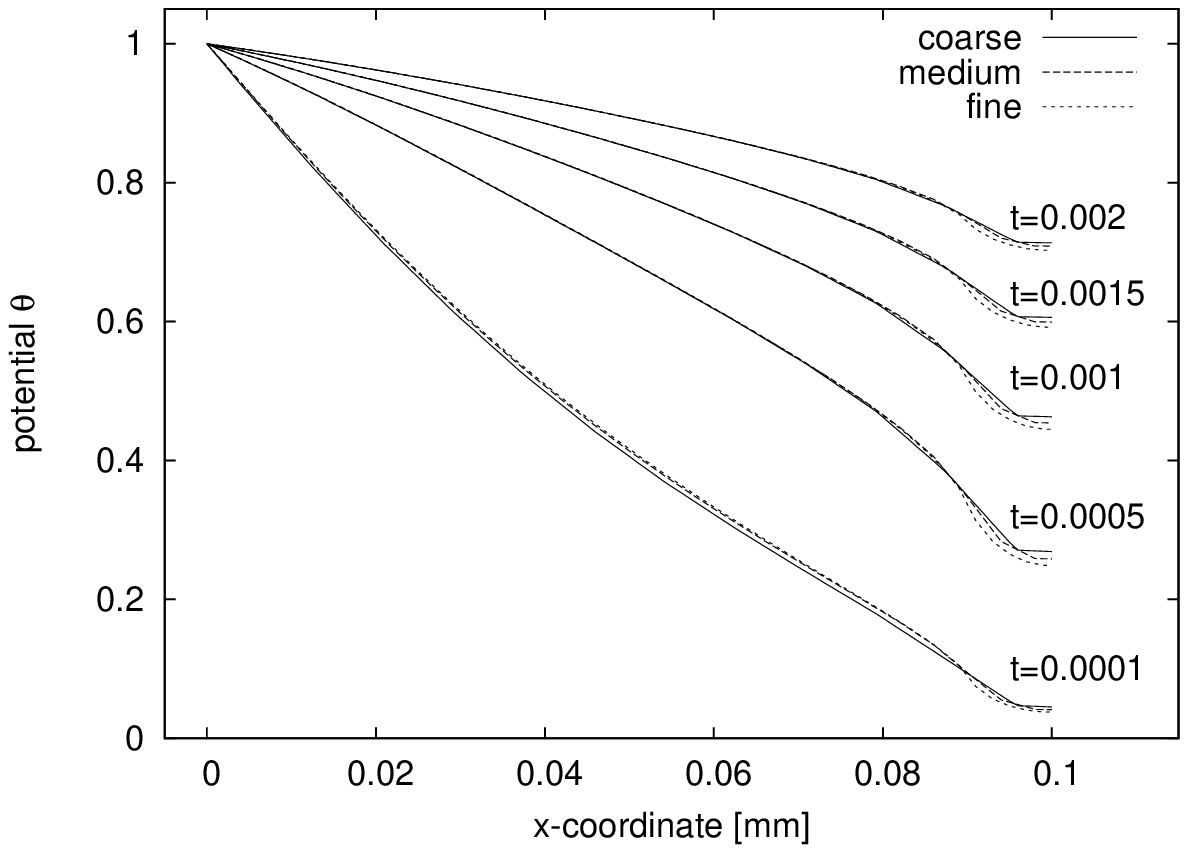,width = 9cm}
\end{center}
\caption{Potential in $x$-direction along the crack at $y=0.05$~m for several time steps.}
\label{fig:xFlow2}
\end{figure}

\begin{figure}
\begin{center}
\epsfig{file=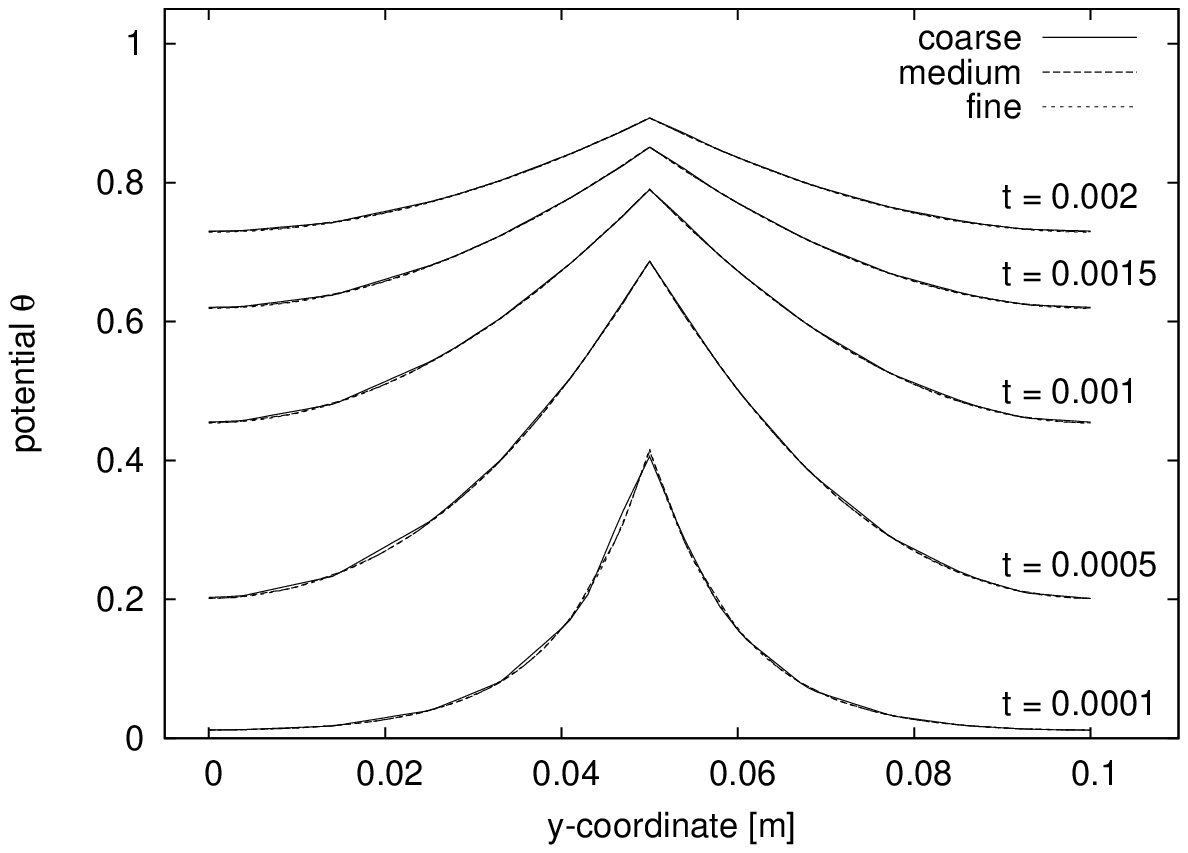,width = 9cm}
\end{center}
\caption{Potential in $y$-direction perpendicular to the crack at $x=0.05$~m for several time steps.}
\label{fig:yFlow2}
\end{figure}

The second example involves the coupling of fracture and nonstationary flow for an aligned lattice. 
The analysis is divided into two steps.
In the first step, a square specimen of length $L=0.1$~m was subjected to an eccentrically applied tensile force $F$. The eccentricity was chosen as $L/4$ with respect to the centerline of the model domain (Fig.~\ref{fig:mesh2}a).
The parameters for the mechanical model were set to $E=40$~GPa, $\gamma = 0.33$, $f_{\rm t}=4$~MPa, $G_{\rm ft} = 100$~N/m, $q=2$, $c=10$, $G_{\rm fc}=50000$~N/m.
The material parameters result in a mechanical response, which is typical for concrete.
Three mechanical lattices with minimum nodal distances of $d_{\rm min} = 0.008$,~0.004~and~0.002~mm were chosen, with the medium density mesh shown in Fig.~\ref{fig:mesh2}a. 
The mesh was aligned in the middle of the specimen, so that the lattice elements were perpendicular to the crack path.
The damage-plasticity model was used only for elements crossing the predefined crack path. All other elements were assumed to be elastic.
This allows one to evaluate the flow potential along the crack.
The result in the form of the load $F$ versus the displacement $d$ at the loading point is shown in Fig.~\ref{fig:ld2}a for three lattices.
The load-displacement curves are almost completely independent of the size of the lattice elements.
At a displacement of $d = 0.1$~m (marked in Fig.~\ref{fig:ld2} by an open circle), the crack almost reaches the right side of the specimen.
For this stage, nonstationary flow analyses were performed for the three lattices dual to the ones used for the mechanical analyses. The lattice with the medium element size is shown in Fig.~\ref{fig:mesh2}b. 
The nodes on the left hand side of the specimen were subjected to $\theta=1$ at all times, whereas the other nodes have an initial potential of $\theta = 0$ at $t=0$. 
The diffusivity $\alpha$ was chosen according to Eq.~(\ref{eq:crackDiff}) with $\alpha_0 = 1$~m$^2$/s and $\varepsilon_{\rm fk} = 0.0025$. 
The results are presented by means of the potential $\theta$ in $x$-direction along the crack path ($y=0.05$~m) in Fig.~\ref{fig:xFlow2}~and the potential $\theta$ in $y$-direction perpendicular to the crack at $x=0.05$~m in Fig.~\ref{fig:yFlow2}. Five time steps of $t = 0.0001$, $0.0005$ , $0.001$, $0.0015$ and $0.002$~s are presented.
The potential along the crack is almost independent of the mesh-size, which is achieved by the special definition of  $\alpha_{\rm c}$ in Eq.~(\ref{eq:crackDiff}).
Perpendicular to the crack, the potential has its maximum at the crack and decreases away from the crack (Fig.~\ref{fig:yFlow2}). 
This imbibition process is described independently of the mesh size. 

\subsection{Fracture and flow in a random mesh}
\begin{figure}
\begin{center}
\begin{tabular}{cc}
\epsfig{file=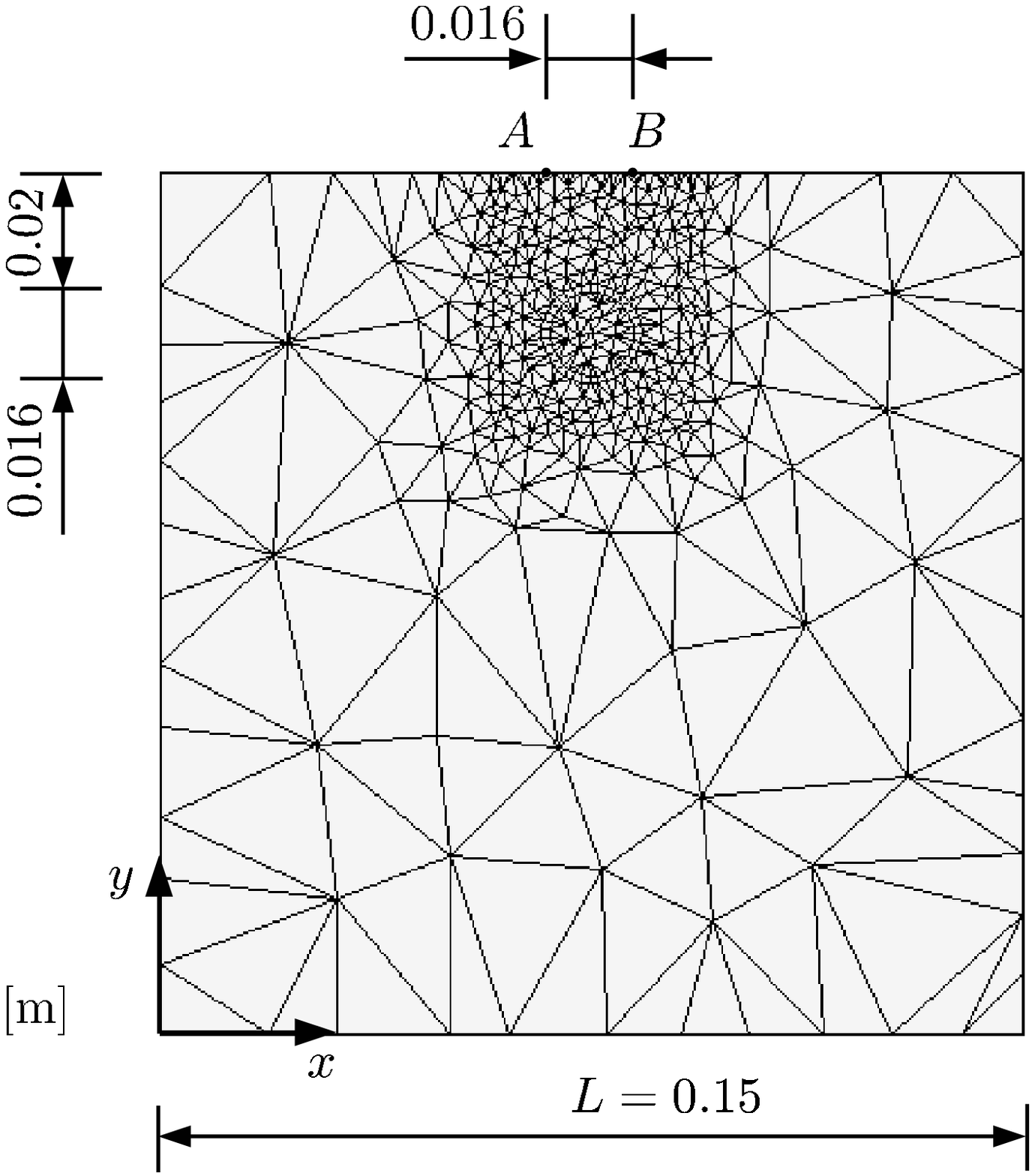,width = 6cm} & \epsfig{file=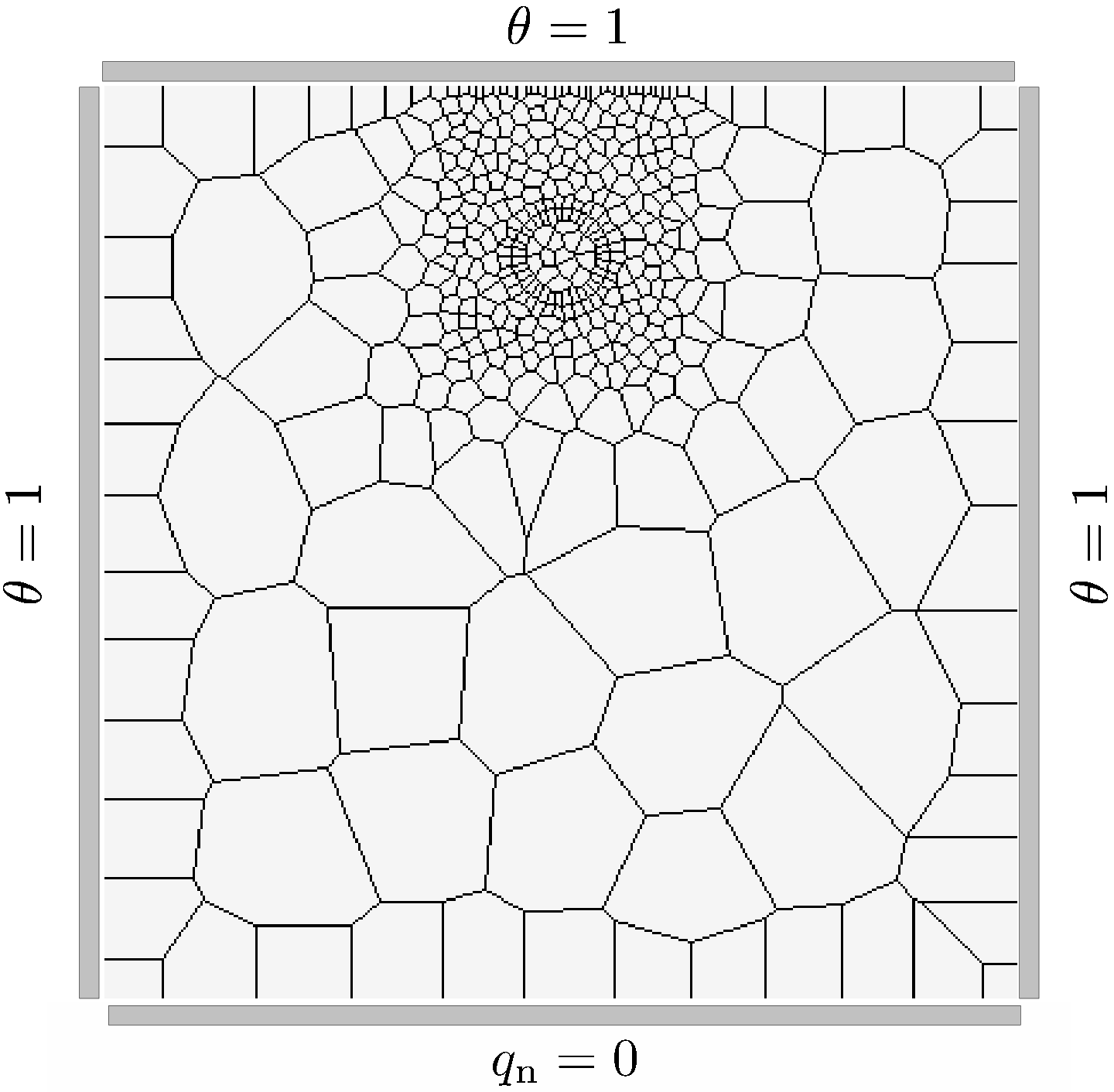,width = 6cm}\\
(a) & (b)
\end{tabular}
\end{center}
\caption{Random lattices with inclusions for (a) structural and (b) transport analysis for $d_{\rm m} = 0.002$~m in the refined regions of the specimens.}
\label{fig:mesh3}
\end{figure}

\begin{figure}
\begin{center}
\epsfig{file=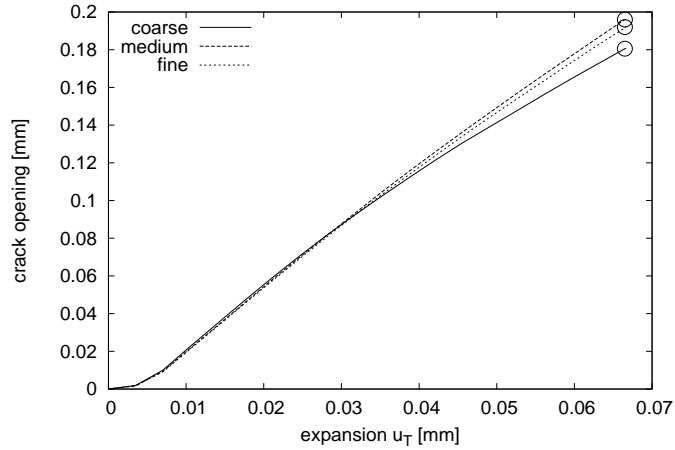,width = 9cm}
\end{center}
\caption{Results of the structural analysis: Crack opening versus eigendisplacement $u_{\rm T}$ for three mesh sizes. The open circles mark stages at which the flow analysis is performed.}
\label{fig:ld3}
\end{figure}

The last example involves crack propagation and flow in a random mesh.
In the previous example the elements in the middle of the specimen were aligned along the crack path, which was useful to evaluate the potential along and perpendicular to the crack.
However, lattices for the analysis of fracture processes are usually random, since regular arrangements of lattice elements influence the direction of crack propagation \cite{JirBaz95}.
In the present example, the mechanical lattice model is used to analyse splitting cracks due to expansion of an inclusion for three random meshes with varying element sizes. 
The specimen geometry and loading setup was chosen according to corrosion experiments in \cite{AndAloMol93}, i.e. the circular inclusion corresponds to a steel reinforcing bar cross-section.
The coarse mesh for structural analysis is shown in Fig.~\ref{fig:mesh3}a.
Within the square specimen of length $L=0.15$~m, an inclusion of diameter $d_a = 16$~mm is located with its center at $x=0.075$~m and $y=0.12$~m.
The expansion of the inclusion is described by eigen-displacement $u_{\rm T}$ subjected to elements, which cross the boundary of the inclusion. 
In these elements the normal component of the eigenstrain in Eq.~(\ref{eq:totStressStrain}) is $\varepsilon_{\rm T} = u_{\rm T}/h$.
Thus, the eigenstrain depends on the lattice element size, whereas the displacement is independent.
A displacement of $u_{\rm T} = 0.0665$~mm at these elements is applied incrementally.
The eigenstrain is applied uniformly along the circumference of the reinforcement bar.
However, the total strain of the lattice elements crossing the circumference of the reinforcement bar is nonuniform, since the mechanical part of the strain depends on the stiffness of the surrounding material.
The crack opening was evaluated as the relative displacement in x-direction of points $A$ and $B$ at the top of the specimen over a length $l_{\rm c} = 0.016$~m (Fig.~\ref{fig:mesh3}a).
The model parameters were the same as for the previous example.
In the refined area, the three lattices were generated with $d_{\rm m} = 0.002$, 0.001 and 0.0005~m.
The results are presented in the form of inclusion expansion $u_{\rm T}$ versus the crack opening in Figure~\ref{fig:ld3}.
The initial part of the expansion-crack opening curve is nearly mesh-independent. 
In a later stage, the results for the three meshes differ.
However, the difference does not appear to be due the element size, since the medium mesh overestimates the crack opening obtained from the fine and coarse mesh.

\begin{figure}
\begin{center}
\epsfig{file=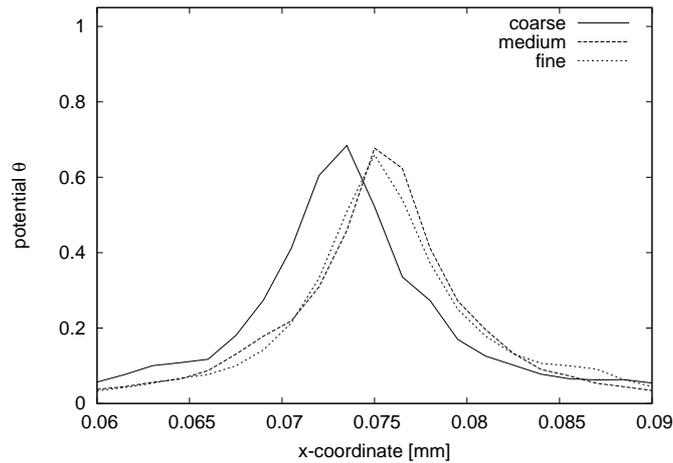,width = 9cm}
\end{center}
\caption{Results of the flow analysis: Potential $\theta$ along the $x$-direction ($y=0.135$~m) in the refined region of the mesh for three element sizes at a eigendisplacement of $u_{\rm T} = 0.0665$~mm and $t = 0.00001$~s.}
\label{fig:xFlow3}
\end{figure}

In the second step of this example, a flow analysis was performed. 
The nodes at the top, left and right boundaries were subjected to a constant potential of $\theta=1$.
All other nodes within the specimen had an initial value of $\theta=0$.  
The diffusivity $\alpha$ was determined according to Eq.~(\ref{eq:diffSplit}).
The results of the potential along the $x$-direction for $y=0.135$~m is shown in Figure~\ref{fig:xFlow3} for $t = 0.00001$~s.
Similarly, as in the second example, the potential has its maximum at the location of the crack, which is not positioned at the same $x$-coordinate since the random lattices are used.
Nevertheless, the shape and magnitude of the potentials along this section of the specimen are independent of the element size.

\section{Conclusions}
In the present work, a lattice approach to model flow in cracked media is presented.
The approach couples mechanical loading to flow analysis by relating the diffusivity of conduit elements to the equivalent crack opening of mechanical lattice elements. Voronoi and Delaunay tessellations are used to define element connectivities for the flow and mechanical lattices, respectively.
The work resulted in the following conclusions:
\begin{itemize}
\item The lattice of conduit elements with cross-sections obtained from the Delaunay triangulation results in an accurate description of stationary flow fields for uncracked homogenous materials.
\item The proposed coupling of mechanical loading with flow analysis results in a mesh-independent description of load-displacement curves and flow fields for lattice elements aligned aligned along cracks.
\item For random lattices, the position of cracks depends on the arrangement of lattice elements. However, the crack openings obtained are independent of the lattice size. Furthermore, flow fields for the cracked material can be described mesh-independently.
\end{itemize}
The present lattice model is capable of describing the coupling of fracture and flow in concrete independent of the element size. The interaction of discrete and continuous flow, which can be problematic in hybrid discrete continuum approaches, is circumvented by modelling both flows by means of the same set of lattice elements, in which the discrete flow is smeared out over the width of the elements.
Future work will concern the extension of this lattice approach to 3D and its application to the modelling of corrosion induced deterioration.

\bibliographystyle{plainnat}

\bibliography{general}

\end{document}